\documentclass[12pt]{article}
\usepackage{amsfonts}
\usepackage{amsmath}
\usepackage{amssymb}
\usepackage{graphicx}
\usepackage{color}
\usepackage[all, knot]{xy}
\usepackage{tikz}

\usepackage{ulem}

\usepackage[utf8]{inputenc}
\usepackage{epstopdf}
\usepackage[footnotesize]{caption}
\usepackage{amsthm}
\usepackage{enumitem}
\usepackage{mathrsfs}

\usepackage[margin=3cm]{geometry}

\def \be {\begin{equation}}
\def \ee {\end{equation}}
\def \bea {\begin{eqnarray}}
\def \eea {\end{eqnarray}}
\def \nn {\nonumber}

\def \rr {\raise.35ex\hbox{\small $\prime$}\kern-.17em{\mbox{\large $\imath$}}}

\def \dels {\partial\kern-.6em /\kern.1em}
\def \As {{A\kern-.5em / \kern.5em}}
\def \Ds {D\kern-.7em / \kern.5em}
\def \a {\alpha}

\def \ks {k\kern-.5em /}
\def \ls {l\kern-.5em /}

\def \s {\sigma}

\newcommand{\ci}[1]{}

\newcommand{\ba}{\begin{eqnarray}}
\newcommand{\ea}{\end{eqnarray}}
\newcommand{\bal}{\begin{align}}
\newcommand{\eal}{\end{align}}
\newcommand{\bay}[1]{\left(\begin{array}{#1}}
\newcommand{\eay}{\end{array}\right)}

\def\DJ{{\fontencoding{T1}\selectfont\char208}}

\newcommand{\hide}[1]{}

\newlist{axioms}{enumerate}{2}
\setlist[axioms,1]{label=\textbf{A\arabic{axiomsi}.}, ref=A\arabic{axiomsi}}
\setlist[axioms,2]{label=\textbf{A\arabic{axiomsi}\rlap{\myEnumCounter{axiomsii}}.},%
                   ref=A\arabic{axiomsi}\myEnumCounter{axiomsii},%
                   align=parleft,%
                   leftmargin=0em,%
                   itemsep=1.4ex,%
                   before={\stepcounter{axiomsi}}}

  \usetikzlibrary{decorations.markings}

\begin{document}

\begin{titlepage}
\begin{center}

\textbf{\LARGE
Early-Time and Late-Time Quantum Chaos
\vskip.3cm
}
\vskip .5in
{\large
Chen-Te Ma$^{a,b,c}$ \footnote{e-mail address: yefgst@gmail.com}}
\\
\vskip 1mm
{\sl
$^a$
Guangdong Provincial Key Laboratory of Nuclear Science,\\
 Institute of Quantum Matter,
South China Normal University, Guangzhou 510006, Guangdong, China.
\\
$^b$
School of Physics and Telecommunication Engineering,\\ 
South China Normal University, Guangzhou 510006, Guangdong, China.
\\
$^c$
The Laboratory for Quantum Gravity and Strings,\\
Department of Mathematics and Applied Mathematics,
University of Cape Town, Private Bag, Rondebosch 7700, South Africa.
}\\
\vskip 1mm
\vspace{40pt}
\end{center}

\begin{abstract}
We show the relation between the Heisenberg averaging of regularized 2-point out-of-time ordered correlation function and the 2-point spectral form factor in bosonic quantum mechanics. The generalization to all even-point is also discussed. We also do the direct extension from the bosonic quantum mechanics to the non-interacting scalar field theory. Finally, we find that the coherent state and large-$N$ approaches are useful in the late-time study. We find that the computation of the coherent state can be simplified by the Heisenberg averaging. Therefore, this provides a simplified way to probe the late-time quantum chaos through a coherent state. The large-$N$ result is also comparable to the $N=3$ numerical result in the large-$N$ quantum mechanics. This can justify that large-$N$ technique in bosonic quantum mechanics can probe the late time, not the early time. Because the quantitative behavior of large-$N$ can be captured from the $N=3$ numerical result, the realization in experiments should be possible.
\end{abstract}
\end{titlepage}

\section{Introduction}
\label{sec:1}
\noindent
{\it Quantum chaos} is an interesting direction in understanding the classical chaos phenomena in a quantum system \cite{Haake:2010}. The classical chaos phenomena have irregular dynamics and instability features. The most intuitive and well-known feature is the sensitivity of initial conditions. Since the well-known condition can be included in other features, this is not a conclusive characteristic, but it is easier to realize in a quantum system like the out-of-time ordered correlation function (OTOC) \cite{Larkin:1969}. This provides the early-time chaos. The other expected criteria are random matrix statistics in the spectrum \cite{Guhr:1997ve} and was concretely realized from the Sinai billiard model \cite{Bohigas:1983er}. Since the spectral statistics in a quantum chaotic model is quite different from an integrable model, this should not be just a coincidence \cite{Evers:2008zz}. Now people conjectured that the spectrum of a quantized chaotic system should be exhibited as in the random matrix ensemble \cite{Ho:2017nyc}. This provides a probe to late-time chaos.
\\

\noindent
Now we introduce early-time chaos. The criteria of early-time chaos are the exponential growth on the time in OTOC. This is characterized by the exponent of OTOC or Lyapunov exponent. In various chaotic models, the OTOC only has the exponential growth at the early time and saturates (vanishing Lyapunov exponent) at the late time. The Sachdev-Ye-Kitaev (SYK) model \cite{Kitaev:2015} is one familiar example realizing early-time chaos \cite{Polchinski:2016xgd}. The physical reason is possibly due to that a quantum system follows the uncertainty principle \cite{Gharibyan:2019sag}. Therefore, we lose the {\it infinitesimal perturbation}. Hence the exponential growth cannot persist forever in a chaotic system. This immediately implies that OTOC cannot probe the classical chaos phenomena \cite{Berry:1979in}. Therefore, we should choose one way to connect OTOC to other late-time chaotic quantities for the study of sensitivity on initial conditions at the late time. 
\\

\noindent
The most direct way for observing the random matrix spectrum is to calculate the level spacing distribution function. The SYK model exhibits the random matrix ensemble \cite{Cotler:2016fpe}. By combining an integrable model with a non-integrable model, one can observe the transition between the Poisson and random matrix distributions \cite{Garcia-Garcia:2017bkg}. This showed that the level spacing distribution function is a useful chaos quantity, but this only restricts to quantum mechanics. The recent study in the random matrix spectrum was extensively studied in the spectral form factor (SFF) \cite{Dyer:2016pou}. Because SFF is defined by the partition function, one can calculate the SFF in quantum field theory without only restricting to quantum mechanics. This lets us probe the random matrix spectrum in two-dimensional conformal field theory (CFT$_2$) \cite{Dyer:2016pou} or other quantum field theories. 
\\

\noindent
Recently, quantum chaos got much attention because the OTOC can be applied to gravity theory or black hole physics. For an analytical calculation in quantum field theory, people chose a regularized form to define the OTOC (regularized OTOC) \cite{Stanford:2015owe}. Ones first showed that the Lyapunov exponent in the regularized OTOC has the universal bound under some assumptions like unitary \cite{Maldacena:2015waa}. Because various studies showed that quantum field theory cannot saturate the bound of the Lyapunov exponent if the bulk theory is not Einstein gravity theory, ones conjectured that the boundary theory of Einstein gravity theory should saturate the bound \cite{Maldacena:2015waa}. The SYK model gave concrete evidence to the conjecture because the model saturates the bound under the holographic limit from a direct computation \cite{Polchinski:2016xgd}, and the corresponding bulk theory can be obtained by the compactification from Einstein gravity theory. This motivates various studies for the conjecture.
\\

\noindent
For understanding the validity of the conjecture, we should know whether the Lyapunov exponent is affected by a {\it regularization}. The first proof for the universal Lyapunov exponent, defined by some regularized forms, for all theories \cite{Tsuji:2017fxs}, but the result is not valid for the generic form. The first issue occurs in the disordered electron system from a practical calculation \cite{Lee:1985zzc}. In the disordered metals \cite{Kamenev:1999zza}, ones first found that the Lyapunov exponent is affected by the regularization \cite{Liao:2018uxa} through the Schwinger-Keldysh formalism \cite{Chamon:1999zz} and non-linear sigma model \cite{Kamenev:2009jj}. Later a more generic study also got a similar result, which showed that the dependence of regularization is due to the infrared regulator \cite{Romero-Bermudez:2019vej}. Nevertheless, the gapless limit possibly does not suffer from the issue \cite{Romero-Bermudez:2019vej}. Hence the conjecture did not meet any counterexample so far.
\\

\noindent
For the probe of black hole physics or information loss \cite{Papadodimas:2015xma}, the first study is the 2-point correlation function in CFT$_2$ \cite{Maldacena:2001kr}. The correlation function decays exponentially with a fast speed, which exhibits that the corresponding back hole is more thermal than a thermal state in a unitary theory. This interpretation links the information loss problem to correlation functions \cite{Almheiri:2013hfa}.
\\

\noindent
The application is not only restricted to high energy physics, and this can also be applied to strongly correlated condensed matter physics \cite{Ma:2018efs} and quantum information. For example, the OTOC provides the butterfly velocity, which characterizes the spread of quantum information \cite{Roberts:2016wdl}. Furthermore, the protocol was devised to measure the OTOC \cite{Swingle:2016var} and was already implemented \cite{Garttner:2016mqj}. The measure of regularized OTOC at a finite temperature was also proposed similarly \cite{Yao:2016ayk}. The protocol can also extract the butterfly velocity from the OTOC \cite{Li:2017pbq}. This experiment set-up can also be applied to the Jaynes-Cummings (JC) interactions \cite{Zhu:2016uws} and the Loschmidt echo \cite{Yan:2020wkt}, which provides a semi-classical calculation of the Lyapunov exponent \cite{Kurchan:2016nju}. These applications provide usefulness to the study of quantum chaos.
\\

\noindent
Recently, the connection between the early-time chaos and late-time chaos has some new developments. One development is using the Lyapunov spectrum to obtain the statistics of the random matrix ensemble. This can be seen as applying the statistics to Lyapunov exponent for connecting the early-time chaos to late-time chaos. Therefore, the {\it averaging} should be an important ingredient for the connection \cite{Gharibyan:2019sag}. The other approach is to average overall Pauli operators in OTOCs to provide the spectral form factors \cite{Cotler:2017jue}. This approach has another similar numerical observation to support from the summing over all momentum modes in OTOC, which provides a similar integrable-chaotic transition to the spectral analysis \cite{Fortes:2019frf}. The first approach only has numerical evidence now. Although the second approach is exact, this is only useful in a {\it finite-dimensional} Hilbert space. 
\\

\noindent
In this paper, we apply the Heisenberg averaging to the regularized 2-point OTOC for obtaining the 2-point SFF \cite{deMelloKoch:2019rxr}. This approach is primarily based on the Haar measure property \cite{Weingarten:1977ya}. We found that applying the Heisenberg averaging to the most generic regularized 4-point OTOC does not give a simple relation to the partition function. However, we can find the corresponding 4-point correlation function with the Heisenberg averaging for getting the four-points SFF. We also discuss the generalization of all even-point. Then we directly extend bosonic quantum mechanics to non-interacting scalar field theory from the Heisenberg group \cite{deMelloKoch:2019rxr}. The extension is not trivial because the dimensions of Hilbert space are infinite. When we consider a Lie algebra with infinite-dimensional generators, we should expect to average over an infinite-dimensional group. For example, $W_{\infty}$ algebra \cite{Bakas:1990sh}. Our result explicitly shows that the averaging is only over a phase space \cite{Zhuang:2019jyq}, not a full Hilbert space. 
\\

\noindent
Since correlation functions and SFFs are hard to compute, the connection seems to be unuseful. Indeed, introducing the Heisenberg averaging to the correlation functions becomes easier to compute at a late-time limit \cite{deMelloKoch:2019rxr}. We first discuss the coherent state \cite{Hepp:1974vg}, which is a quantum state closest to a classical regime. Using the coherent state \cite{Hepp:1974vg} can simplify calculation in the averaged OTOCs because the calculation can be simplified by using the property of the Heisenberg group. In a generic system, we can also use the saddle-point evaluation to treat the integration in the OTOCs \cite{deMelloKoch:2019rxr}. Our coherent approach is also suitable to the solvable and detectable model, the 2-photon non-degenerate JC model with a rotating wave approximation, which ignores the oscillating fast term \cite{Iwasawa:1995}. The second approach is the large-$N$ technique \cite{Itzykson:1979fi}. We apply the large-$N$ approximation to bosonic quantum mechanics \cite{Chowdhury:2017jzb}. The large-$N$ theory is just harmonic oscillators with a modified frequency. Therefore, 2-point SFF has an exact solution. We compare this exact solution to a numerical study. Our $N=3$ numerical result is already good enough for obtaining a quantitative behavior of the large-$N$. This precisely justifies that the large-$N$ technique in bosonic systems \cite{Weinstein:2005kw} is useful for the probe of late time.
\\

\noindent
This paper is organized as follows. We first introduce the Heisenberg averaging to correlation functions for connecting to SFFs in Sec.~\ref{sec:2}. Then we discuss the late-time study from the coherent state approach and large-$N$ technique in Sec.~\ref{sec:3}. Finally, we discuss and conclude in Sec.~\ref{sec:4}. 

\section{Heisenberg Group}
\label{sec:2}
\noindent
We first introduce the Heisenberg averaging to the regularized 2-point OTOC at a finite temperature \cite{Larkin:1969}, and it leads to the 2-point SFF \cite{Dyer:2016pou}. Then we discuss the generalization to higher-point correlation functions and also scalar field theory \cite{deMelloKoch:2019rxr}.

\subsection{2-Point SFF}
\noindent
The Heisenberg group is a two-dimensional Lie group generated by the canonical position $X$ and the canonical momentum $P$. The commutation between $X$ and $P$ is 
\bea
\lbrack P, X\rbrack=-i.
\eea
 An element of this group with respect to the variables, $q_1$, $q_2$, is 
\bea
U(q_1,q_2)\equiv e^{iq_1 X+i q_2 P}.
\eea
This element satisfies the below property 
\bea
U(q_1,q_2)U^{\dagger}(q_1,q_2) =1.
\eea
\\

\noindent
By a direct calculation:
\bea
&&\int_{-\infty}^{\infty} \frac{dq_1}{2\pi}\int_{-\infty}^{\infty}dq_2\ \langle x_1|U(q_1, q_2)|x_2\rangle\langle y_1|U^{\dagger}(q_1, q_2)|y_2\rangle
\nn\\
&=&
\int_{-\infty}^{\infty}\frac{dq_1}{2\pi}\int_{-\infty}^{\infty}dq_2\ e^{iq_1x_2-iq_1y_1}
\nn\\
&&\times\langle x_1|e^{iq_2P}|x_2\rangle\langle y_1|e^{-iq_2P}|y_2\rangle
\nn\\
&=&
\int_{-\infty}^{\infty}\frac{dq_1}{2\pi}\int_{-\infty}^{\infty}dq_2\ e^{iq_1x_2-iq_1y_1}\langle x_1| x_2-q_2\rangle\langle y_1|y_2+q_2\rangle
\nn\\
&=&
\int_{-\infty}^{\infty}dq_2\ \delta(x_2-y_1)\langle x_1|x_2-q_2\rangle\langle x_2|y_2+q_2\rangle
\nn\\
&=&\int_{-\infty}^{\infty}dq_2\ \delta(x_2-y_1)\delta(x_1-x_2+q_2)\delta(x_2-y_2-q_2)
\nn\\
&=&\delta(x_2-y_1)\delta(x_2-y_2+x_1-x_2)
\nn\\
&=&\delta(x_2-y_1)\delta(x_1-y_2),
\eea
in which we used:
\bea
U(q_1, q_2)&\equiv& e^{iq_1X+iq_2P}=e^{iq_2P}e^{iq_1X}e^{\frac{-iq_1q_2}{2}};
\nn\\
e^{iq_1X}|x\rangle&=&e^{iq_1x}|x\rangle
\eea
in the first equality, we used:
\bea
e^{iq_2P}|x\rangle=|x-q_2\rangle; \qquad \langle p|x\rangle=e^{-ipx}
\eea
in the second equality, we used
\bea
\int_{-\infty}^{\infty}\frac{dq}{2\pi}\ e^{iq_1(x_2-y_1)}=\delta(x_2-y_1)
\eea
in the third equality. The result shows the same result as in using the Haar measure  \cite{Weingarten:1977ya}.
\\

\noindent
Now we calculate the Heisenberg average of the regularized 2-point OTOC:
\bea
&&C_{2}
\nn\\
&\equiv&\int_{-\infty}^{\infty} \frac{dq_1}{2\pi}\int_{-\infty}^{\infty}dq_2\int_{-\infty}^{\infty} dx\ 
\nn\\
&&\times
\langle x| U(q_1, q_2)e^{-(\beta/2+it)H}U^{\dagger}(q_1, q_2)e^{-(\beta/2-it)H}|x\rangle
\nn\\
&=&\int_{-\infty}^{\infty} \frac{dq_1}{2\pi}\int_{-\infty}^{\infty}dq_2\int_{-\infty}^{\infty} dx\int_{-\infty}^{\infty}dx_1\int_{-\infty}^{\infty}dx_2\int_{-\infty}^{\infty}dx_3\ 
\nn\\
&&\times\langle x| U(q_1, q_2)|x_1\rangle\langle x_1| e^{-(\beta/2+it)H}|x_2\rangle
\nn\\
&&\times\langle x_2|U^{\dagger}(q_1, q_2)|x_3\rangle
\langle x_3|e^{-(\beta/2-it)H}|x\rangle
\nn\\
&=&\int_{-\infty}^{\infty} dx\int_{-\infty}^{\infty}dx_1\int_{-\infty}^{\infty}dx_2\int_{-\infty}^{\infty}dx_3\ 
\nn\\
&&\times\delta(x-x_3)\delta(x_1-x_2)
\nn\\
&&\times\langle x_1| e^{-(\beta/2+it)H}|x_2\rangle\langle x_3|e^{-(\beta/2-it)H}|x\rangle
\nn\\
&=&\int_{-\infty}^{\infty}dx\int_{-\infty}^{\infty}dx_1\
\nn\\
&&\times \langle x_1| e^{-(\beta/2+it)H}|x_1\rangle\langle x|e^{-(\beta/2-it)H}|x\rangle.
\eea
Therefore, the 2-point SFF at the inverse finite temperature $\beta/2$ exactly matches $C_2$ at the inverse temperature $\beta$.

\subsection{Higher-Point Correlation Functions}
\noindent
Here we consider different generalizations to study the relation between correlation functions and spectral form factors.
We first consider the most generic Heisenberg average of the regularized 4-point OTOC \cite{Romero-Bermudez:2019vej}
\bea
&&C_{\mathrm{M}4}
\nn\\
&\equiv&\int_{-\infty}^{\infty}\frac{dq_1}{2\pi}\int^{\infty}_{-\infty}dq_2\int_{-\infty}^{\infty}\frac{dq_3}{2\pi}\int_{-\infty}^{\infty}dq_4\int_{-\infty}^{\infty}dx\ 
\nn\\
&&\times
\bigg\langle x\bigg| U_1(q_1, q_2) e^{-(\beta\sigma+it)H}U_1^{\dagger}(q_1, q_2)
\nn\\
&&\times e^{-\big(\beta(\alpha-\sigma)-it\big)H}
U_2(q_3, q_4) e^{-(\beta\sigma+it)H}
\nn\\
&&\times
U_2^{\dagger}(q_3, q_4)e^{-\big(\beta(1-\alpha-\sigma)-it\big)H}
\bigg|x\bigg\rangle
\nn\\
&&+\int_{-\infty}^{\infty}\frac{dq_1}{2\pi}\int^{\infty}_{-\infty}dq_2\int_{-\infty}^{\infty}\frac{dq_3}{2\pi}\int_{-\infty}^{\infty}dq_4\int_{-\infty}^{\infty}dx\ 
\nn\\
&&\times
\bigg\langle x\bigg| e^{-\big(\beta(1-\alpha-\sigma)+it\big)H}
U_2(q_3, q_4) e^{-(\beta\sigma-it)H}
\nn\\
&&\times U_2^{\dagger}(q_3, q_4)e^{-\big(\beta(\alpha-\sigma)+it\big)H}
\nn\\
&&\times
U_1(q_1, q_2) e^{-(\beta\sigma-it)H}
U_1^{\dagger}(q_1, q_2)
\bigg|x\bigg\rangle,
\eea
where
\bea
0\le\alpha\le 1; \qquad 0\le\sigma\le\frac{1}{4}.
\eea
\\

\noindent
We will use the following relation
\bea
&&\int^{\infty}_{-\infty}\frac{dq_1}{2\pi}\int^{\infty}_{-\infty}dq_2\ \langle x_1|U(q_1, q_2)|x_2\rangle\langle y_1|U^{\dagger}(q_1, q_2)|y_2\rangle
\nn\\
&=&\delta(x_2-y_1)\delta(x_1-y_2)
\eea
to do the below integration:
\bea
&&\int^{\infty}_{-\infty}\frac{dq_1}{2\pi}\int^{\infty}_{-\infty}dq_2\ 
\nn\\
&&\times
\bigg\langle x\bigg|U_1(q_1, q_2)e^{-(\beta\sigma+it)H}U_1^{\dagger}(q_1, q_2)\bigg| y_1\bigg\rangle
\nn\\
&=&\delta(x-y_1)\bigg\langle x\bigg|e^{(-\beta\sigma+it)H}\bigg| y_1\bigg\rangle,
\nn\\
&&\int^{\infty}_{-\infty} \frac{dq_3}{2\pi}\int^{\infty}_{-\infty}dq_4\ 
\nn\\
&&\times\bigg\langle y_2\bigg|U_2(q_3, q_4)e^{-(\beta\sigma+it)H}U_2^{\dagger}(q_3, q_4)\bigg| y_3\bigg\rangle
\nn\\
&=&\delta(y_2-y_3)\bigg\langle y_2\bigg|e^{(-\beta\sigma+it)H}\bigg| y_3\bigg\rangle.
\eea
Therefore, we get: 
\bea
&&\bigg\langle x\bigg| U_1(q_1, q_2) e^{-(\beta\sigma+it)H}U_1^{\dagger}(q_1, q_2)e^{-\big(\beta(\alpha-\sigma)-it\big)H}
\nn\\
&&\times
U_2(q_3, q_4) e^{-(\beta\sigma+it)H}
U_2^{\dagger}(q_3, q_4)e^{-\big(\beta(1-\alpha-\sigma)-it\big)H}
\bigg|x\bigg\rangle
\nn\\
&=&
\int^{\infty}_{-\infty}dy_1\int^{\infty}_{-\infty}dy_2\int^{\infty}_{-\infty}dy_3\
\delta(x-y_1)\delta(y_2-y_3)
\nn\\
&&
\times
\bigg\langle x\bigg|e^{(-\beta\sigma+it)H}\bigg| y_1\bigg\rangle
\bigg\langle y_1\bigg|e^{-\big(\beta(\alpha-\sigma)-it\big)H}\bigg| y_2\bigg\rangle
\nn\\
&&
\times
\bigg\langle y_2\bigg|e^{(-\beta\sigma+it)H}\bigg| y_3\bigg\rangle
\bigg\langle y_3\bigg|e^{-\big(\beta(1-\alpha-\sigma)-it\big)H}\bigg| x\bigg\rangle
\nn\\
&=&
\int^{\infty}_{-\infty}dy_2\
\bigg\langle x\bigg|e^{(-\beta\sigma+it)H}\bigg| x\bigg\rangle
\bigg\langle x\bigg|e^{-\big(\beta(\alpha-\sigma)-it\big)H}\bigg| y_2\bigg\rangle
\nn\\
&&
\times
\bigg\langle y_2\bigg|e^{(-\beta\sigma+it)H}\bigg| y_2\bigg\rangle
\bigg\langle y_2\bigg|e^{-\big(\beta(1-\alpha-\sigma)-it\big)H}\bigg| x\bigg\rangle.
\eea
Finally, we obtain
\bea
&&C_{\mathrm{M}4}
\nn\\
&=&\int^{\infty}_{-\infty}dx\int^{\infty}_{-\infty}dy_2\
\nn\\
&&\times
\bigg\langle x\bigg|e^{(-\beta\sigma+it)H}\bigg| x\bigg\rangle
\bigg\langle x\bigg|e^{-\big(\beta(\alpha-\sigma)-it\big)H}\bigg| y_2\bigg\rangle
\nn\\
&&
\times
\bigg\langle y_2\bigg|e^{(-\beta\sigma-it)H}\bigg| y_2\bigg\rangle
\bigg\langle y_2\bigg|e^{-\big(\beta(1-\alpha-\sigma)+it\big)H}\bigg| x\bigg\rangle
\nn\\
&&+
\int^{\infty}_{-\infty}dx\int^{\infty}_{-\infty}dy_2\
\nn\\
&&\times
\bigg\langle x\bigg|e^{(-\beta\sigma-it)H}\bigg| x\bigg\rangle
\bigg\langle y_2\bigg|e^{-\big(\beta(\alpha-\sigma)+it\big)H}\bigg| x\bigg\rangle
\nn\\
&&
\times
\bigg\langle y_2\bigg|e^{(-\beta\sigma+it)H}\bigg| y_2\bigg\rangle
\nn\\
&&\times
\bigg\langle x\bigg|e^{-\big(\beta(1-\alpha-\sigma)-it\big)H}\bigg| y_2\bigg\rangle.
\eea
\\

\noindent
When we choose:
\bea
\alpha=\frac{1}{2}; \qquad \sigma=\frac{1}{4},
\eea
we can go back to the usual regularized 4-point OTOC, and the regularized  4-point OTOC becomes
\bea
&&C_{\mathrm{M}4}
\nn\\
&=&\int^{\infty}_{-\infty}dx\int^{\infty}_{-\infty}dy_2\
\nn\\
&&\times
\bigg\langle x\bigg|e^{(-\beta/4+it)H}\bigg| x\bigg\rangle
\bigg\langle x\bigg|e^{-(\beta/4-it)H}\bigg| y_2\bigg\rangle
\nn\\
&&
\times
\bigg\langle y_2\bigg|e^{(-\beta/4+it)H}\bigg| y_2\bigg\rangle
\bigg\langle y_2\bigg|e^{-(\beta/4-it)H}\bigg| x\bigg\rangle
\nn\\
&&
+\int^{\infty}_{-\infty}dx\int^{\infty}_{-\infty}dy_2\
\nn\\
&&\times
\bigg\langle x\bigg|e^{(-\beta/4-it)H}\bigg| x\bigg\rangle
\bigg\langle y_2\bigg|e^{-(\beta/4+it)H}\bigg| x\bigg\rangle
\nn\\
&&
\times
\bigg\langle y_2\bigg|e^{(-\beta/4-it)H}\bigg| y_2\bigg\rangle
\bigg\langle x\bigg|e^{-(\beta/4+it)H}\bigg| y_2\bigg\rangle.
\eea
It is hard to find a simple relation to the partition function. 
\\

\noindent
It is not surprising for this result. For this method, we need the relation in the 2-point SFF
\bea
A_1B_1=I,
\eea
where $A_1$ and $B_1$ are two operators, and $I$ is the identity operator. Therefore, we only have one independent operator. We used the Heisenberg group to generate the operator. If we want to generalize the relation to the 4-point SFF, we need the below relation
\bea
A_1B_1A_2B_2=I.
\eea
Therefore, we need to generate three independent variables, and the operator $B_2$ is 
\bea
B_2=A_2^{\dagger}B_1^{\dagger}A_1^{\dagger}.
\eea
\\

\noindent
One direct way for obtaining the 4-point SFF is to use the Heisenberg group to generate three independent operators, and then we can work the similar generalization \cite{Cotler:2017jue} by calculating the 4-point correlation function
\bea
U_1(q_1, q_2)U_2(q_3, q_4; t)U_3(q_5, q_6)U_4(q_1, q_2, q_3, q_4, q_5, q_6; t),
\nn\\
\eea
where
\bea
&&U_2(q_3 ,q_4; t)
\nn\\
&\equiv& e^{iHt}U_2(q_3, q_4)e^{-iHt};
\nn\\
&&U_4(q_1, q_2, q_3, q_4, q_5, q_6; t)
\nn\\
&\equiv& e^{iHt}U_4(q_1, q_2, q_3, q_4, q_5, q_6)e^{-iHt};
\nn\\
&&U_4(q_1, q_2, q_3, q_4, q_5, q_6)
\nn\\
&\equiv& U_3^{\dagger}(q_5, q_6)U_2^{\dagger}(q_3, q_4)U_1^{\dagger}(q_1, q_2)
\eea
with the Heisenberg averaging.
For the generalization to all even-point OTOCs, we need to introduce the condition
\bea
A_1B_1A_2B_2\cdots A_kB_k=I.
\eea
The $U_4$ have more variables than $U_1$, $U_2$, and $U_3$ because we want to have the below condition:
\bea
&&U_1(q_1, q_2)U_2(q_3, q_4)U_3(q_5, q_6)U_4(q_1, q_2, q_3, q_4, q_5, q_6)
\nn\\
&=&U_1(q_1, q_2)U_2(q_3, q_4)U_3(q_5, q_6)U_3^{\dagger}(q_5, q_6)U_2^{\dagger}(q_3, q_4)U_1^{\dagger}(q_1, q_2)
\nn\\
&=&I.
\eea
\\

\noindent
After we do the $(2k-1)$-times Heisenberg averaging, we can get the $2k$-point SFFs. This generalization is more similar to $2k$-point correlation functions, not our familiar OTOCs, which at most has two independent operators. The physical interpretation of the higher-point correlation functions possibly not be the same as the OTOCs, but it is interesting to study the decay of correlation functions for the information loss issue \cite{Maldacena:2001kr}. Although the generalization was only discussed at $\beta=0$, it is easy to use a similar way to generalize to $\beta\ne 0$ as in our previous study (2-point SFF).
 
\subsection{Scalar Field Theory}
\noindent
Here we discuss how to extend the above result to quantum field theory. Although bosonic quantum mechanics has an infinite-dimensional Hilbert space, it is still different from the quantum field theory. One way for touching this problem is to consider the scalar field theory in a box, which satisfies the periodic boundary condition, for having the discrete momenta $\vec{k}$. 
The Hilbert space is also countable, and spectrum is also discrete. Therefore, the scalar field theory in a box should have a similar result to the bosonic quantum mechanics. 
This gives a more direct and the first approach.
\\

\noindent
Here we consider the non-interacting scalar field theory. This can be seen as the assembly of harmonic oscillators. Because a direct generalization is easy from the discrete momenta, we use the oscillator languages:
\bea
a=\frac{(P-i\omega X)}{\sqrt{2\omega}};\qquad  a^\dagger =\frac{(P+i\omega X)}{\sqrt{2\omega}}
\eea
to rewrite $X$ and $P$ in terms of the creation operator $a$ and annihilation operator $a^{\dagger}$. The Heisenberg group are same, but it is written in terms of $a$ and $a^{\dagger}$:
\bea
&&U\lbrack q_1(\cdot),q_2(\cdot)\rbrack
\nn\\
&=&
e^{\sum_{\vec k}a_{\vec k}\left(iq_2(\vec k)\sqrt{\omega_{\vec k}\over 2}
-{q_1(\vec k)\over \sqrt{2\omega_{\vec k}}}\right)}
e^{\sum_{\vec p}a^\dagger_{\vec p} \left(iq_2(\vec p)\sqrt{\omega_{\vec p}\over 2}
+{q_1(\vec p)\over \sqrt{2\omega_{\vec p}}}\right)}
\nn\\
&&\times
e^{\sum_{\vec l}\left( {q_1^2(\vec l)\over 4\omega}+{q_2^2(\vec l)\omega\over 4}\right)}.
\eea
The square bracket notation is just to stress that $U$ is the functional of $q_1(\vec k)$ and $q_2(\vec k)$. This Heisenberg group in quantum field theory is totally generalized from the below:
\bea
U(q_1,q_2)
&=&e^{iq_1X+iq_2P}
\nn\\
&=&e^{a\left(iq_2\sqrt{\omega\over 2}-{q_1\over \sqrt{2\omega}}\right)}
e^{a^\dagger \left(iq_2\sqrt{\omega\over 2}+{q_1\over \sqrt{2\omega}}\right)}
e^{{q_1^2\over 4\omega}+{q_2^2\omega\over 4}}.
\nn\\
\eea
\\

\noindent
The non-interacting scalar field theory in a box also has the same Hamiltonian form as in the harmonic oscillator
\bea
H_{\mathrm{NS}}=\frac{1}{V}\sum_{\vec{k}}\ \frac{1}{2} \tilde{a}^{\dagger}(\vec{k})\tilde{a}(\vec{k}),
\eea
where $V$ is the volume of the box. The $\tilde{a}^{\dagger}(\vec{k})$ and $\tilde{a}(\vec{k})$ are the standard creation and annihilation operators in the box, and then they satisfy the commutation relation:
\bea
\lbrack \tilde{a}(\vec{k}_1), \tilde{a}^{\dagger}(\vec{k}_2)\rbrack&=&2V\omega_{\vec{k}_1}\delta_{\vec{k}_1\vec{k}_2};
\nn\\
\lbrack \tilde{a}(\vec{k}_1), \tilde{a}(\vec{k}_2)\rbrack&=&0;
\nn\\
\lbrack \tilde{a}^{\dagger}(\vec{k}_1), \tilde{a}^{\dagger}(\vec{k}_2)\rbrack&=&0, 
\eea
 where
\bea
\omega_{\vec{k}_1}^2\equiv |\vec{k}_1|^2+m^2
\eea
 with the mass of the scalar field $m$. Performing the field redefinition
\bea
\tilde{a}(\vec{k})\equiv \sqrt{2V\omega(\vec{k})}a(\vec{k})
\eea
can exactly see the assembly of harmonic oscillators. Hence the non-interacting scalar field theory in a box must have the same relation between the Heisenberg average of correlation functions and SFFs as in bosonic quantum mechanics.

\section{Late-Time Study}
\label{sec:3}
\noindent
Since SFFs \cite{Dyer:2016pou} and OTOTs \cite{Larkin:1969} are hard to compute, the connection seems to be unuseful practically. However, this is not true exactly. What we are interested in is the late-time physics. A late-time limit is also a classical limit, which is useful for the probe of classical chaos. Therefore, the coherent state is one approach. We demonstrate that the two-particles coherent state can simplify a calculation of the Heisenberg average of OTOCs \cite{deMelloKoch:2019rxr}. The other computable approach is a large-$N$ study. We use the large-$N$ bosonic quantum mechanics to present. The large-$N$ theory is the harmonic oscillators with a modified frequency. Since the large-$N$ quantum chromodynamics (QCD) and various large-$N$ quantum field theory can approach the non-interacting quantum field theory, our result can be applied to a large-$N$ quantum field theory also \cite{deMelloKoch:2019rxr}. Finally, we compare an exact solution of the 2-point SFF to a numerical solution. For the $N=3$ numerical solution, we already obtain a quantitative comparison to the large-$N$. 

\subsection{Coherent State}
\noindent
We first demonstrate why the Heisenberg averaging can simply the calculation of coherent state from the example, 2-particle coherent state :
\bea
 a_1|\alpha_1\alpha_2\rangle =\alpha_1|\alpha_1\alpha_2\rangle,\qquad
 a_2|\alpha_1\alpha_2\rangle =\alpha_2|\alpha_1\alpha_2\rangle;  
\nn
\eea
\bea
 |\alpha_1\alpha_2\rangle
=e^{-\frac{|\alpha_1|^2+|\alpha_2|^2}{2}}e^{\alpha_1 a_1^\dagger+\alpha_2 a_2^\dagger}|0,0\rangle.
\eea
The completeness relation of this coherent state is
\bea
\int {d^2\alpha_1\over \pi}\int {d^2\alpha_2\over \pi}\
|\alpha_1\a_2\rangle\langle\alpha_1\alpha_2|=1.
\eea
Because we have two particles (two canonical pairs), we have four variables ($q_1$, $q_2$, $r_1$, $r_2$) in the Heisenberg group
\bea
U(q_1,q_2,r_1,r_2)=e^{iq_1 X_1+iq_2 P_1+ir_1 X_2+ir_2P_2}.
\eea
We can find that the computation of the regularized 2-point OTOC
\bea
\label{C(t)}
&&C_2(t)
\nn\\
&=&\langle \alpha_1\alpha_2 |U(q_1,q_2,r_1,r_2)e^{-\frac{\beta H}{2} -iHt}
\nn\\
&&\times
 U^{\dagger}(q_1,q_2,r_1,r_2)
 e^{-\frac{\beta H}{2}+i Ht}|\alpha_1\alpha_2\rangle
\eea
always meets the matrix element $\langle \alpha_1\alpha_2 |U(q_1,q_2,r_1,r_2)|\gamma_1\gamma_2\rangle$. Because the matrix elements related to the coherent state parameters are just Gaussian, and the exponent of Heisenberg group only has the linear term on $a$ and $a^{\dagger}$, this implies that the integration over the coherent variables are simplified, which entirely relies on the property of the Heisenberg group. If we choose other operators, the Gaussian form can be broken. Therefore, we think that the Heisenberg averaging is a nice simplification for a coherent state. This coherent study can be applied to the observable and exactly solvable model, the 2-photon non-degenerate JC model with a rotating wave approximation \cite{Iwasawa:1995}. Because this model is solvable, we can carry out all integration exactly. In general, we need to do a saddle-point evaluation on the integration at the late time. Here we only discussed the 2-particle coherent state, but it is easy to do the similar generalization to other particle numbers.

\subsection{Large-$N$ Quantum Mechanics}
\noindent
Now we discuss the second approach for late-time physics. The large-$N$ limit gives the factorization to simplify a study \cite{Itzykson:1979fi}. We demonstrate this approach by the large-$N$ bosonic quantum mechanics
\bea
H_{\mathrm{QMN}}=\sum_{j=1}^N \frac{P^j P^j}{2}+ \mu^2\frac{X^j X^j}{2}+g\frac{(X^j X^j)^2}{4},
\eea
where $j=1,2,\cdots,N$, and $g$ is the coupling constant. We first introduce the auxiliary field $\sigma$, and consider the large-$N$ approximation ($\sigma$ is just a constant) to obtain
\bea
H_{\mathrm{QMNM}}=\sum_{j=1}^N\frac{P^j P^j}{2}+\mu^2\frac{X^j X^j}{2}+\lambda\sigma \frac{X^j X^j}{2},
\eea
in which the ’t Hooft coupling constant $\lambda \equiv gN$ is fixed when we scale $N\to\infty$. Because the $\sigma$ is just a constant under the large-$N$ limit, this theory is just harmonic oscillators with a modified frequency. We determine the auxiliary field
\bea
\sigma = \frac{\sum_{j=1}^N\langle X^j X^j\rangle}{N}
\eea
from the two-point function or the large-$N$ Schwinger-Dyson equation
\bea
\left({d^2\over dt^2}+\mu^2 +\lambda\sigma\right) \sum_{j=1}^N\langle X^j(t)X^j(t')\rangle =-iN\delta (t-t').
\nn\\
\eea
Therefore, we obtain the solution
\bea
\sigma =\frac{1}{2\sqrt{\mu^2+\lambda\sigma}},
\eea
by taking $t-t^{\prime}=\epsilon$ and choosing $\epsilon\to 0^+$ from the below calculation:
\bea
&&\langle X^j(t)X^j(t')\rangle
\nn\\
&=&\int {d\omega\over 2\pi}{iN\over \omega^2-\mu^2-\lambda\s +i\epsilon}e^{i\omega (t-t')}
\nn\\
&=&{N\theta (t-t')\over 2\sqrt{\mu^2+\lambda\sigma}}e^{-i\sqrt{\mu^2+\lambda\sigma}\, (t-t')}
\nn\\
&&
+{N\theta (t'-t)\over 2\sqrt{\mu^2+\lambda\sigma}}e^{i\sqrt{\mu^2+\lambda\sigma}\, (t-t')}\,.
\eea
\\

\begin{figure}[h]
\begin{centering}
\includegraphics[width=0.32\textwidth]{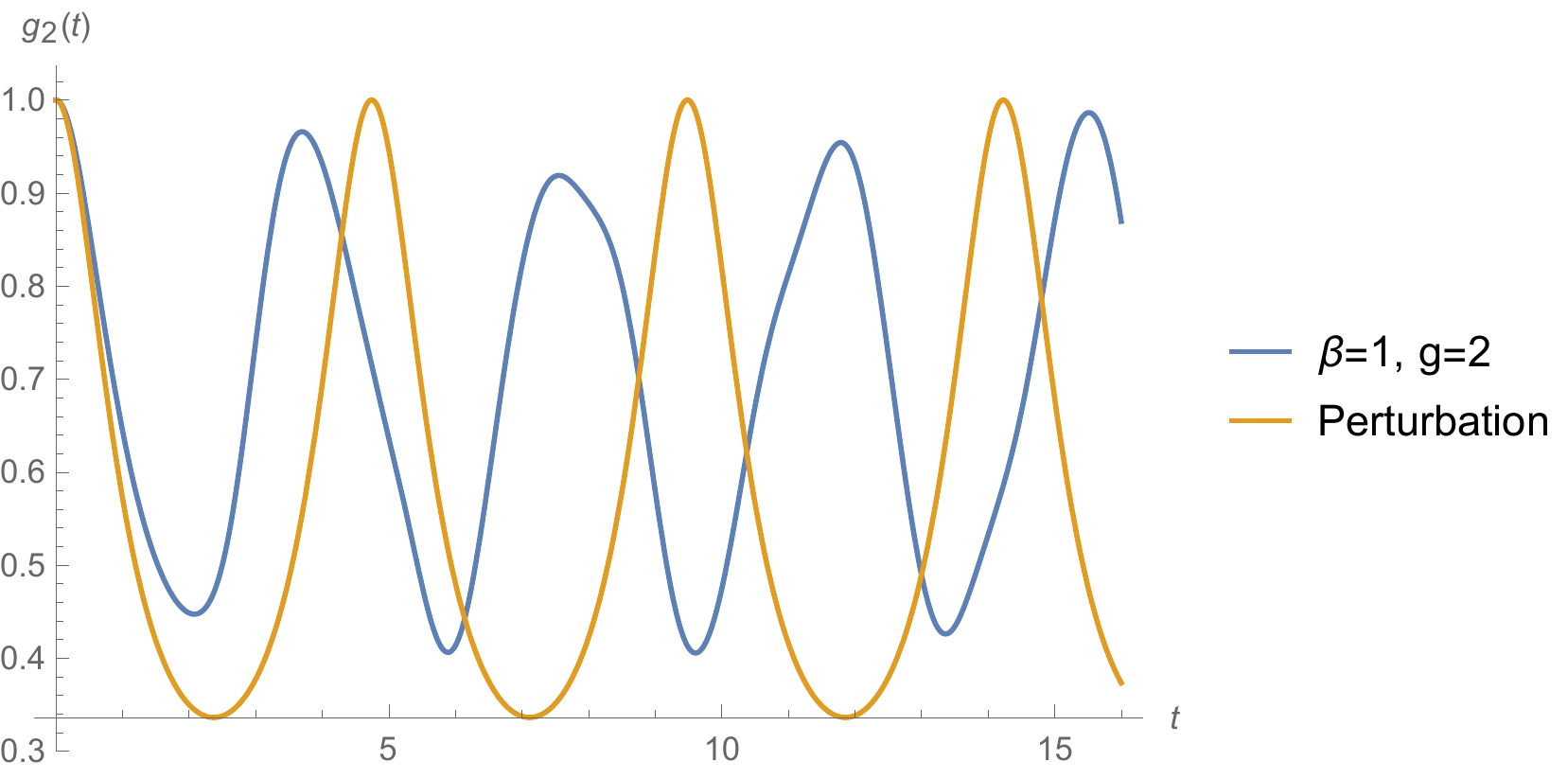}
\hfill
\includegraphics[width=0.32\textwidth]{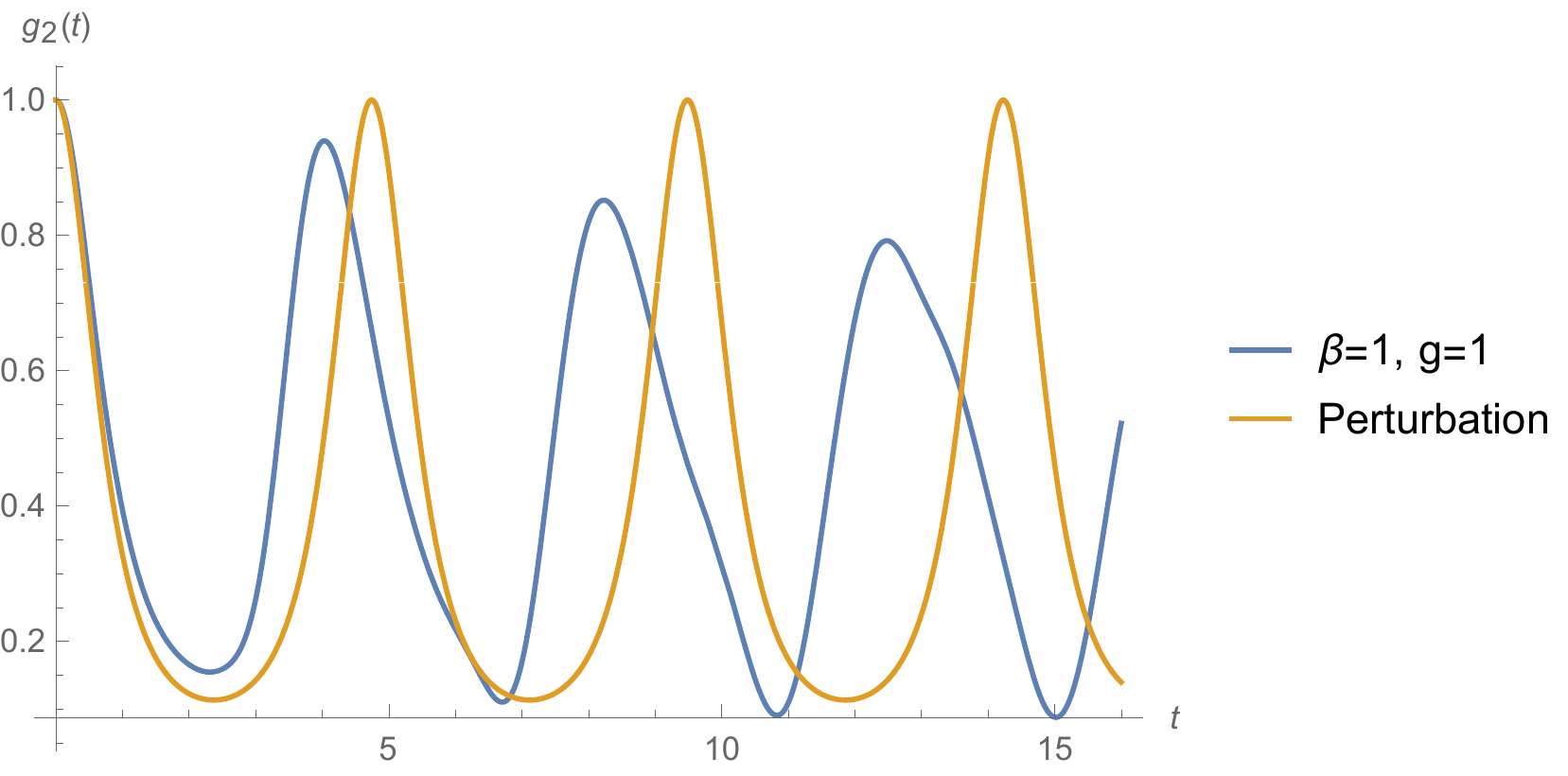}
\hfill
\includegraphics[width=0.32\textwidth]{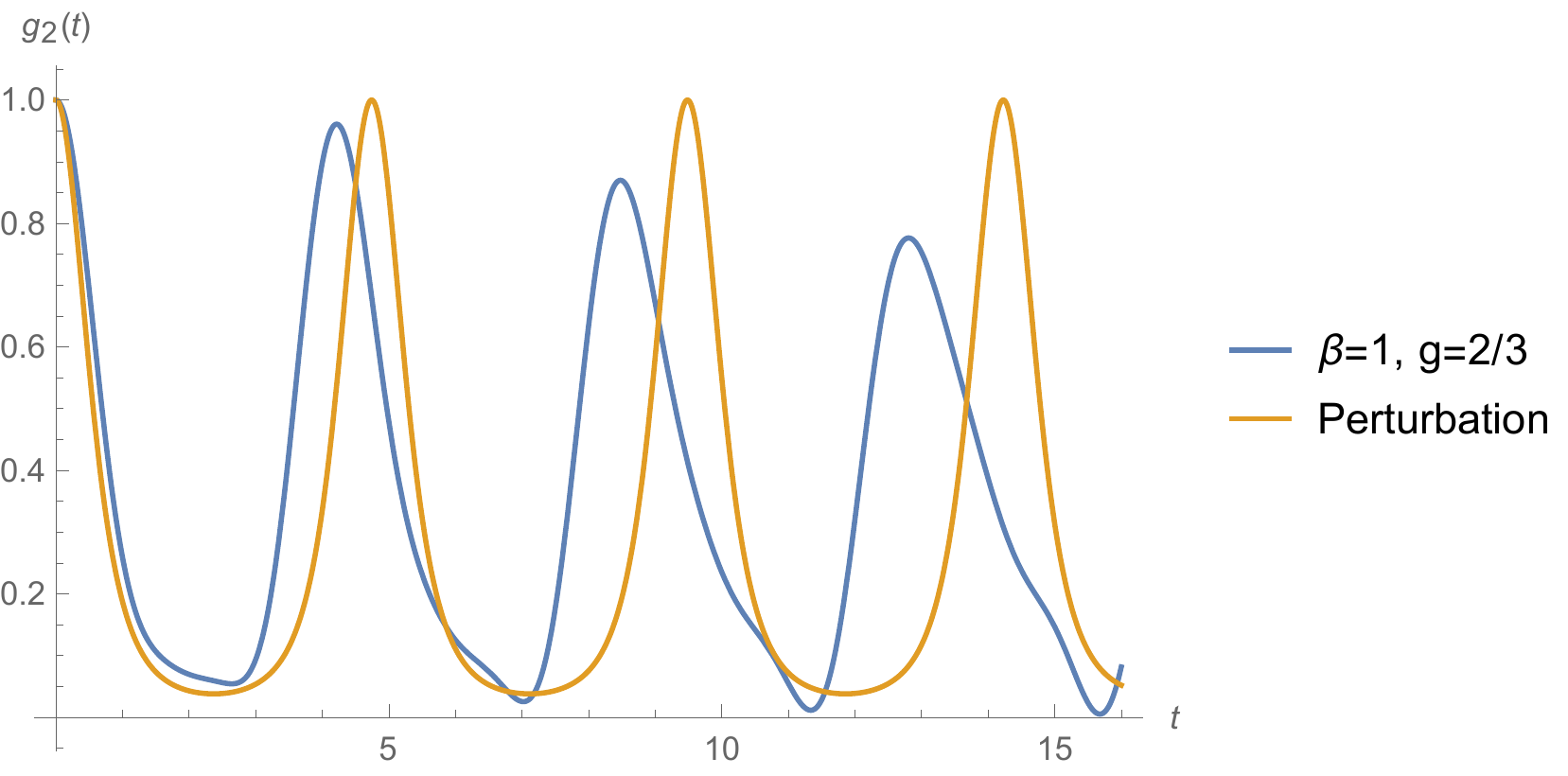}
\par\end{centering}
\caption{We fix the inverse temperature $\beta=1$ while choosing the ’t Hooft coupling constant $\lambda$=2. The lattice size is 8 in $N$=1, and the lattice size is 4 in $N$=2, 3. The number of lattice points is 128 in $N$=1, and the number of lattice points is 32 in $N$=2, 3. We calculate the 2-point SFF $g_2(t)$ from 16 low-lying eigenenergy modes for $N$=1, 2, and 3 in the left, middle, and right figures respectively. The numerical solution in $N$=3 matches the large-$N$ perturbation quantitatively for $t=0-3$. 
Although the deviation becomes larger for a longer time, the large $N$ decreases the deviation.  
Hence the large-$N$ limit is equivalent to the classical or late-time limits.
}
\label{123.pdf}
\end{figure}

\noindent
The 2-point SFF 
\bea
g_2(\beta, t)
&\equiv&\bigg|\frac{\sum_{n_1=0}^{\infty} {\cal D}(n_1) e^{(-\beta+it)E_{n_1}}}{\sum_{n_2=0}^{\infty} {\cal D}(n_2) e^{-\beta E_{n_2}}}\bigg|^2
\eea
can be calculated exactly \cite{deMelloKoch:2019rxr}:
\bea
&&\sum_{n=0}^{\infty} {\cal D}(n) e^{(-\beta+it)E_{n}}
\nn\\
&=&\sum_{n=0}^{\infty} \frac{(n+N-1)!}{n!(N-1)!} e^{(-\beta+it)(\frac{\omega N}{2}+n\omega)}
\nn\\
&=&e^{(-\beta+it)\frac{\omega N}{2}}\sum_{n=0}^{\infty}\frac{(n+N-1)!}{n!(N-1)!} e^{n(-\beta+it)\omega}
\nn\\
&=&e^{(-\beta+it)\frac{\omega N}{2}}\big(1-e^{(-\beta+it)\omega}\big)^{-N},
\eea
where 
\bea
\omega\equiv\sqrt{\mu^2+\lambda\sigma}, \qquad E_n\equiv\frac{\omega N}{2}+n\omega;
\nn
\eea
\bea
\sum_{n=0}^{\infty}\frac{(n+N-1)!}{n!(N-1)!} x^n=(1-x)^{-N}, \qquad |x|<1;
\eea
\bea
g_2(\beta, t)
&\equiv&\bigg|\frac{\sum_{n_1=0}^{\infty} {\cal D}(n_1) e^{(-\beta+it)E_{n_1}}}{\sum_{n_2=0}^{\infty} {\cal D}(n_2) e^{-\beta E_{n_2}}}\bigg|^2
\nn\\
&=&
\bigg(\frac{1+e^{-2\omega\beta}
-2e^{-\omega\beta}}
{1+e^{-2\omega\beta}
-2\cos (\omega t)e^{-\omega\beta}}\bigg)^N
\eea
for $\beta\ne 0$. When we take $\beta=0$, the two-points SFF is divergent. Therefore, we are careful about an ordering of the summation. We choose the same ordering of the summation as in a numerical study that we will work. Therefore, this should give the rigorously consistent result even for the extremely-low $\beta$ for the comparison between the exact solution and the numerical solution. Our numerical result for $N=1$, $N=2$, and $N=3$ with a fixed inverse temperature $\beta=1$ and ’t Hooft coupling constant $\lambda$=2 are given in Fig. \ref{123.pdf}. 
\\

\noindent
We use the naive discretization to treat the derivative term or the momenta
\bea
P_j^2\psi_j\equiv-\frac{\psi_{j+1}-2\psi_j+\psi_{j-1}}{a^2},
\eea
where $P_j$ is the lattice momentum, $\psi_j$ is the lattice eigenfunction, and $a$ is the lattice spacing, in the Hamiltonian. The lattice index is labeled by $j=1, 2, \cdots, n$, where $n$ is the number of lattice points. Then we do an exact diagonalization to obtain eigenvalues of the Hamiltonian. If the lattice size 
\bea
2L=n\cdot a
\eea 
is not large enough, the high-energy modes will not provide a bound state in the numerical study. Therefore, we always choose some numbers in the low-lying eigenenegy modes for obtaining the physically correct result. Here our numerical study always preserves the periodic boundary condition:
\bea
&&-L\le X_j<L; \qquad X_1=-L;
\nn\\
 &&X_{j+1}\equiv X_j+a; \qquad X_{n+1}\equiv X_1.
\eea
We use the subscript in $X$ to denote the lattice index without confusing to the superscript in $X$, which denotes the number of the canonical positions.
\\

\noindent
 The comparison justifies that increasing $N$ in the numerical solution helps the probe of late time physics. It is also surprising that the $N=3$ numerical result is enough to capture the quantitative large-$N$ result for $t=0-3$. This sheds the light in the hope of the experimental realization on the large-$N$ physics \cite{Swingle:2016var} \cite{Yao:2016ayk}. Although the deviation between the perturbation and numerical result becomes larger, letting $N$ be larger can reduce the deviation. 
 Hence the large-$N$ perturbation approaches the classical or late-time limits.

\section{Discussion and Conclusion}
\label{sec:4}
\noindent
We used the Heisenberg averaging to connect the regularized 2-point OTOC \cite{Larkin:1969} and the 2-point SFF in any bosonic quantum mechanics \cite{deMelloKoch:2019rxr}. The connection used the average of phase space, not a full Hilbert space. This avoids the naive intuition,  averaging infinite-dimensional variables. This is an exact realization for connecting the early-time chaos to late-time chaos in an infinite-dimensional Hilbert space. For obtaining the generalization to the higher-point SFFs, we also found the corresponding higher-point correlation functions with the Heisenberg averaging. The higher-point correlation functions possibly do not have the same physical interpretation as in the OTOCs, but it is still useful to study the decay in correlation functions for the motivation of information loss \cite{Maldacena:2001kr}. Although the bosonic quantum mechanics has the infinite-dimensional Hilbert space, it is still countable. Since the scalar field theory in a box, the spectrum is discrete and the Hilbert space is countable as in the bosonic quantum mechanics. Therefore, we could do a direct generalization from bosonic quantum mechanics to the scalar field theory \cite{deMelloKoch:2019rxr}.
\\

\noindent
 The correlation functions and SFFs \cite{Dyer:2016pou} are hard to compute both. Therefore, the connection is not practical. For the motivation of quantum chaos, we are more interested in the late-time limit. We used the coherent state and the large-$N$ technique to show usefulness at a late-time limit. In the coherent state, the Heisenberg averaging simplifies the computation for all coherent states \cite{deMelloKoch:2019rxr}. In general, the coherent state cannot provide an exact solution, but we can use the saddle-point evaluation to do the perturbation. The large-$N$ theory can have a simplification from a factorization \cite{Itzykson:1979fi}, and the theory is just the non-interacting theory. We demonstrated the large-$N$ study from the large-$N$ bosonic quantum mechanics \cite{deMelloKoch:2019rxr}. This gives harmonic oscillators with a modified frequency \cite{deMelloKoch:2019rxr}. Therefore, the exact solution in the SFFs can be obtained \cite{deMelloKoch:2019rxr}. We compared the exact solution of the 2-point SFF to the numerical solution, This explicitly shows that the large-$N$ technique is really useful for the probe of late-time physics, not early-time physics. We also found that the $N=3$ result already gives the quantitative result to the large-$N$ study for $t=0-3$. The experimental realization is impossible for infinite oscillators, but it is possible for three oscillators. Therefore, the numerical study also confirmed the hope of an experimental realization \cite{Swingle:2016var} \cite{Yao:2016ayk}.
\\

\noindent
The connection between the correlation functions and SFFs is primarily based on the Heisenberg group. The elements of the Heisenberg group can also generate a coherent state. When we computed the coherent states, we also found a useful simplification. The use of the coherent state can be seen as a classical limit. We know that the sensitivity to initial conditions is also hidden in the irregular dynamics and instability in 1-interval classical chaos. Therefore, the connection through the Heisenberg averaging should not be just a coincidence, and this possibly has a more fundamental reason behind it. This direction should be understood more from the practical computation of quantum chaos using a coherent state. Because the connection purely relies on the Heisenberg group, we should study other groups to know the differences. This should be helpful to know why using the Heisenberg group is quite special,
\\

\noindent
We provided a coherent state and the large-$N$ techniques for showing that the connection can be computed practically. The examples in this paper are all solvable and integrable models. Therefore, we did not give an example from chaotic model. However, the demonstrated approaches should be useful for getting the plateau time scale in a non-integrable model from a saddle-point evaluation (coherent state) or large-$N$ perturbation. The speed of the decay in correlation functions is closely related to the information loss issue \cite{Maldacena:2001kr}. Now we can use the coherent state and the large-$N$ technique to compute SFFs for probing information loss issues from quantum chaos perspective. Therefore, calculating SFFs or correlation functions should be interesting based on the connection. Here we only studied large-$N$ theory at leading-order. 
It is equivalent to studying the non-interacting theory. 
The large-$N$ perturbation to the next-order should show the non-integrability from the interaction. 
Because the interaction becomes for the large-$N$, a study of $N=1$ should be most non-integrable. 
Hence studying in detail under the connection should be interesting.    
\\

\noindent
In the final, we want to comment about the connection between the OTOC and SFF from quantum field theory. 
Now our generalization is only for non-interacting scalar field theory because we need to transform the $X$ and $P$ to the creation and annihilation operators. 
When scalar field theory in a box, the summation of momenta is discrete. Hence it is easy to generalize the connection. 
However, including an interacting term usually does not have such a relation. 
Hence we suggest that an application of the adaptive perturbation method \cite{Weinstein:2005kw} to the connection between the OTOC and SFF in scalar field theory. 
In this method, one can use the same transformation from $X$ and $P$ to the creation and annihilation operators \cite{Weinstein:2005kw}. 
Hence using the adaptive perturbation method can obtain the connection in non-integrable quantum field theory. 
In this method, the spectrum can be calculated accurately \cite{Weinstein:2005kw}. 
Hence applying the adaptive perturbation method should be realizable.

\section*{Acknowledgments}
\noindent
The author would like to thank Robert de Mello Koch, Yingfei Gu, Hendrik J.R. Van Zyl, Shinsei Ryu, and Beni Yoshida for their useful discussions.
\\

\noindent
The author was supported by the Post-Doctoral International Exchange Program and China Postdoctoral Science Foundation, Postdoctoral General Funding: Second Class (Grant No.
2019M652926) and would like to thank Nan-Peng Ma for his encouragement.
\\

\noindent
The author would like to thank the Yukawa Institute for Theoretical Physics at the Kyoto University and Institute of Theoretical Physics at the Chinese Academy of Sciences.
\\

\noindent
Discussions during the workshops, ``Quantum Information and String Theory'' and ``Workshop on Holography and Quantum Matter'', were useful to complete this work.



\baselineskip 22pt
\begin{thebibliography}{99}

\bibitem{Haake:2010}
F.~Haake,
``Quantum signatures of chaos,''
Springer-Verlag, Berlin, Heidelberg, 2010.
doi:10.1007/978-3-642-05428-0

\bibitem{Larkin:1969}
  A.~I.~Larkin and Yu.~N.~Ovchinnikov,
  ``Quasiclassical Method in the Theory of Superconductivity,''
  JETP {\bf 28}, 1200 (1969).

\bibitem{Guhr:1997ve}
  T.~Guhr, A.~Muller-Groeling and H.~A.~Weidenmuller,
  ``Random matrix theories in quantum physics: Common concepts,''
  Phys.\ Rept.\  {\bf 299}, 189 (1998)
  doi:10.1016/S0370-1573(97)00088-4
  [cond-mat/9707301].

\bibitem{Bohigas:1983er}
  O.~Bohigas, M.~J.~Giannoni and C.~Schmit,
  ``Characterization of chaotic quantum spectra and universality of level fluctuation laws,''
  Phys.\ Rev.\ Lett.\  {\bf 52}, 1 (1984).
  doi:10.1103/PhysRevLett.52.1

\bibitem{Evers:2008zz}
  F.~Evers and A.~D.~Mirlin,
  ``Anderson transitions,''
  Rev.\ Mod.\ Phys.\  {\bf 80}, 1355 (2008).
  doi:10.1103/RevModPhys.80.1355

\bibitem{Ho:2017nyc}
  W.~W.~Ho and \DJ.~ Radičević
  ``The Ergodicity Landscape of Quantum Theories,''
  Int.\ J.\ Mod.\ Phys.\ A {\bf 33}, no. 04, 1830004 (2018)
  doi:10.1142/S0217751X18300041
  [arXiv:1701.08777 [quant-ph]].

\bibitem{Kitaev:2015}
A.~Kitaev,
``A simple model of quantum holography. - 2015,''
 Talks at KITP,
 April 7 and May 27.

\bibitem{Polchinski:2016xgd}
  J.~Polchinski and V.~Rosenhaus,
  ``The Spectrum in the Sachdev-Ye-Kitaev Model,''
  JHEP {\bf 1604}, 001 (2016)
  doi:10.1007/JHEP04(2016)001
  [arXiv:1601.06768 [hep-th]].

\bibitem{Gharibyan:2019sag}
  H.~Gharibyan, M.~Hanada, B.~Swingle and M.~Tezuka,
  ``A characterization of quantum chaos by two-point correlation functions,''
  arXiv:1902.11086 [quant-ph].

\bibitem{Berry:1979in}
  M.~V.~Berry and N.~L.~Balazs,
  ``Evolution Of Semiclassical Quantum States In Phase Space,''
  J.\ Phys.\ A {\bf 12}, 625 (1979).
  doi:10.1088/0305-4470/12/5/012

\bibitem{Cotler:2016fpe}
  J.~S.~Cotler {\it et al.},
  ``Black Holes and Random Matrices,''
  JHEP {\bf 1705}, 118 (2017)
  Erratum: [JHEP {\bf 1809}, 002 (2018)]
  doi:10.1007/JHEP09(2018)002, 10.1007/JHEP05(2017)118
  [arXiv:1611.04650 [hep-th]].

\bibitem{Garcia-Garcia:2017bkg}
  A.~M.~García-García, B.~Loureiro, A.~Romero-Bermúdez and M.~Tezuka,
  ``Chaotic-Integrable Transition in the Sachdev-Ye-Kitaev Model,''
  Phys.\ Rev.\ Lett.\  {\bf 120}, no. 24, 241603 (2018)
  doi:10.1103/PhysRevLett.120.241603
  [arXiv:1707.02197 [hep-th]].

\bibitem{Dyer:2016pou}
  E.~Dyer and G.~Gur-Ari,
  ``2D CFT Partition Functions at Late Times,''
  JHEP {\bf 1708}, 075 (2017)
  doi:10.1007/JHEP08(2017)075
  [arXiv:1611.04592 [hep-th]].

\bibitem{Stanford:2015owe}
  D.~Stanford,
  ``Many-body chaos at weak coupling,''
  JHEP {\bf 1610}, 009 (2016)
  doi:10.1007/JHEP10(2016)009
  [arXiv:1512.07687 [hep-th]].

\bibitem{Maldacena:2015waa}
  J.~Maldacena, S.~H.~Shenker and D.~Stanford,
  ``A bound on chaos,''
  JHEP {\bf 1608}, 106 (2016)
  doi:10.1007/JHEP08(2016)106
  [arXiv:1503.01409 [hep-th]].

\bibitem{Tsuji:2017fxs}
  N.~Tsuji, T.~Shitara and M.~Ueda,
  ``Bound on the exponential growth rate of out-of-time-ordered correlators,''
  Phys.\ Rev.\ E {\bf 98}, 012216 (2018)
  doi:10.1103/PhysRevE.98.012216
  [arXiv:1706.09160 [cond-mat.stat-mech]].

\bibitem{Lee:1985zzc}
  P.~A.~Lee and T.~V.~Ramakrishnan,
  ``Disordered electronic systems,''
  Rev.\ Mod.\ Phys.\  {\bf 57}, 287 (1985).
  doi:10.1103/RevModPhys.57.287

\bibitem{Kamenev:1999zza}
  A.~Kamenev and A.~Andreev,
  ``Electron-electron interactions in disordered metals: Keldysh formalism,''
  Phys.\ Rev.\ B {\bf 60}, 2218 (1999).
  doi:10.1103/PhysRevB.60.2218

\bibitem{Liao:2018uxa}
  Y.~Liao and V.~Galitski,
  ``Nonlinear sigma model approach to many-body quantum chaos: Regularized and unregularized out-of-time-ordered correlators,''
  Phys.\ Rev.\ B {\bf 98}, no. 20, 205124 (2018)
  doi:10.1103/PhysRevB.98.205124
  [arXiv:1807.09799 [cond-mat.dis-nn]].

\bibitem{Chamon:1999zz}
  C.~Chamon, A.~W.~W.~Ludwig and C.~Nayak,
  ``Schwinger-Keldysh approach to disordered and interacting electron systems: Derivation of Finkelstein's renormalization-group equations,''
  Phys.\ Rev.\ B {\bf 60}, 2239 (1999).
  doi:10.1103/PhysRevB.60.2239

\bibitem{Kamenev:2009jj}
  A.~Kamenev and A.~Levchenko,
  ``Keldysh technique and nonlinear sigma-model: Basic principles and applications,''
  Adv.\ Phys.\  {\bf 58}, 197 (2009)
  doi:10.1080/00018730902850504
  [arXiv:0901.3586 [cond-mat.other]].
  
\bibitem{Romero-Bermudez:2019vej}
A.~Romero-Bermúdez, K.~Schalm and V.~Scopelliti,
``Regularization dependence of the OTOC. Which Lyapunov spectrum is the physical one?,''
JHEP \textbf{07}, 107 (2019)
doi:10.1007/JHEP07(2019)107
[arXiv:1903.09595 [hep-th]].

\bibitem{Papadodimas:2015xma}
  K.~Papadodimas and S.~Raju,
  ``Local Operators in the Eternal Black Hole,''
  Phys.\ Rev.\ Lett.\  {\bf 115}, no. 21, 211601 (2015)
  doi:10.1103/PhysRevLett.115.211601
  [arXiv:1502.06692 [hep-th]].

\bibitem{Maldacena:2001kr} 
  J.~M.~Maldacena,
  ``Eternal black holes in anti-de Sitter,''
  JHEP {\bf 0304}, 021 (2003)
  doi:10.1088/1126-6708/2003/04/021
  [hep-th/0106112].

\bibitem{Almheiri:2013hfa}
  A.~Almheiri, D.~Marolf, J.~Polchinski, D.~Stanford and J.~Sully,
  ``An Apologia for Firewalls,''
  JHEP {\bf 1309}, 018 (2013)
  doi:10.1007/JHEP09(2013)018
  [arXiv:1304.6483 [hep-th]].

\bibitem{Ma:2018efs} 
  C.~T.~Ma,
  ``Parity Anomaly and Duality Web,''
  Fortsch.\ Phys.\  {\bf 66}, no. 8-9, 1800045 (2018)
  doi:10.1002/prop.201800045
  [arXiv:1802.08959 [hep-th]].

\bibitem{Roberts:2016wdl}
  D.~A.~Roberts and B.~Swingle,
  ``Lieb-Robinson Bound and the Butterfly Effect in Quantum Field Theories,''
  Phys.\ Rev.\ Lett.\  {\bf 117}, no. 9, 091602 (2016)
  doi:10.1103/PhysRevLett.117.091602
  [arXiv:1603.09298 [hep-th]].

\bibitem{Swingle:2016var}
  B.~Swingle, G.~Bentsen, M.~Schleier-Smith and P.~Hayden,
  ``Measuring the scrambling of quantum information,''
  Phys.\ Rev.\ A {\bf 94}, no. 4, 040302 (2016)
  doi:10.1103/PhysRevA.94.040302
  [arXiv:1602.06271 [quant-ph]].

\bibitem{Garttner:2016mqj}
  M.~Gärttner, J.~G.~Bohnet, A.~Safavi-Naini, M.~L.~Wall, J.~J.~Bollinger and A.~M.~Rey,
  ``Measuring out-of-time-order correlations and multiple quantum spectra in a trapped ion quantum magnet,''
  Nature Phys.\  {\bf 13}, 781 (2017)
  doi:10.1038/nphys4119
  [arXiv:1608.08938 [quant-ph]].

\bibitem{Yao:2016ayk}
  N.~Y.~Yao, F.~Grusdt, B.~Swingle, M.~D.~Lukin, D.~M.~Stamper-Kurn, J.~E.~Moore and E.~A.~Demler,
  ``Interferometric Approach to Probing Fast Scrambling,''
  arXiv:1607.01801 [quant-ph].

\bibitem{Li:2017pbq}
  J.~Li, R.~Fan, H.~Wang, B.~Ye, B.~Zeng, H.~Zhai, X.~Peng and J.~Du,
  ``Measuring Out-of-Time-Order Correlators on a Nuclear Magnetic Resonance Quantum Simulator,''
  Phys.\ Rev.\ X {\bf 7}, no. 3, 031011 (2017)
  doi:10.1103/PhysRevX.7.031011
  [arXiv:1609.01246 [cond-mat.str-el]].

\bibitem{Zhu:2016uws}
  G.~Zhu, M.~Hafezi and T.~Grover,
  ``Measurement of many-body chaos using a quantum clock,''
  Phys.\ Rev.\ A {\bf 94}, no. 6, 062329 (2016)
  doi:10.1103/PhysRevA.94.062329
  [arXiv:1607.00079 [quant-ph]].

\bibitem{Yan:2020wkt}
B.~Yan, L.~Cincio and W.~H.~Zurek,
``Information Scrambling and Loschmidt Echo,''
Phys. Rev. Lett. \textbf{124}, no.16, 160603 (2020)
doi:10.1103/PhysRevLett.124.160603
[arXiv:1903.02651 [quant-ph]].

\bibitem{Kurchan:2016nju}
  J.~Kurchan,
  ``Quantum bound to chaos and the semiclassical limit,''
  arXiv:1612.01278 [cond-mat.stat-mech].

\bibitem{Cotler:2017jue}
  J.~Cotler, N.~Hunter-Jones, J.~Liu and B.~Yoshida,
  ``Chaos, Complexity, and Random Matrices,''
  JHEP {\bf 1711}, 048 (2017)
  doi:10.1007/JHEP11(2017)048
  [arXiv:1706.05400 [hep-th]].

\bibitem{Fortes:2019frf}
E.~M.~Fortes, I.~García-Mata, R.~A.~Jalabert and D.~A.~Wisniacki,
``Gauging classical and quantum integrability through out-of-time ordered correlators,''
Phys. Rev. E \textbf{100}, no.4, 042201 (2019)
doi:10.1103/PhysRevE.100.042201
[arXiv:1906.07706 [quant-ph]].

\bibitem{deMelloKoch:2019rxr} 
  R.~de Mello Koch, J.~H.~Huang, C.~T.~Ma and H.~J.~R.~Van Zyl,
  ``Spectral Form Factor as an OTOC Averaged over the Heisenberg Group,''
  Phys.\ Lett.\ B {\bf 795}, 183 (2019)
  doi:10.1016/j.physletb.2019.06.025
  [arXiv:1905.10981 [hep-th]].

\bibitem{Weingarten:1977ya}
  D.~Weingarten,
  ``Asymptotic Behavior of Group Integrals in the Limit of Infinite Rank,''
  J.\ Math.\ Phys.\  {\bf 19}, 999 (1978).
  doi:10.1063/1.523807

\bibitem{Bakas:1990sh}
  I.~Bakas and E.~B.~Kiritsis,
  ``Structure and representations of the W(infinity) algebra,''
  Prog.\ Theor.\ Phys.\ Suppl.\  {\bf 102}, 15 (1990).
  doi:10.1143/PTPS.102.15

\bibitem{Zhuang:2019jyq} 
  Q.~Zhuang, T.~Schuster, B.~Yoshida and N.~Y.~Yao,
  ``Scrambling and Complexity in Phase Space,''
  Phys.\ Rev.\ A {\bf 99}, no. 6, 062334 (2019)
  doi:10.1103/PhysRevA.99.062334
  [arXiv:1902.04076 [quant-ph]].

\bibitem{Hepp:1974vg}
  K.~Hepp,
  ``The Classical Limit for Quantum Mechanical Correlation Functions,''
  Commun.\ Math.\ Phys.\  {\bf 35}, 265 (1974).
  doi:10.1007/BF01646348

\bibitem{Iwasawa:1995}
 H.~Iwasawa and K.~Matsuo,
``Exact solutions of Heisenberg operators for two-photon non-degenerate Jaynes-Cummings model,''
    Optics\ communications\ {\bf 117}, 550-559 (1995).
    doi:10.1016/0030-4018(95)00130-Z

\bibitem{Itzykson:1979fi}
  C.~Itzykson and J.~B.~Zuber,
  ``The Planar Approximation. 2.,''
  J.\ Math.\ Phys.\  {\bf 21}, 411 (1980).
  doi:10.1063/1.524438

\bibitem{Chowdhury:2017jzb}
  D.~Chowdhury and B.~Swingle,
  ``Onset of many-body chaos in the $O(N)$ model,''
  Phys.\ Rev.\ D {\bf 96}, no. 6, 065005 (2017)
  doi:10.1103/PhysRevD.96.065005
  [arXiv:1703.02545 [cond-mat.str-el]].
  
\bibitem{Weinstein:2005kw}
M.~Weinstein,
``Adaptive perturbation theory. I. Quantum mechanics,''
[arXiv:hep-th/0510159 [hep-th]].






\end{thebibliography}
\end{document}